## *Phase transitions and thermal-stress-induced structural changes in a ferroelectric $Pb(Zr_{0.80}Ti_{0.20})O_3$ single crystal*

J. Frantti,[1] Y. Fujioka,[1] A. Puretzky,[2] Y. Xie,[3] Z.-G. Ye,[3] C. Parish[2] and A. M. Glazer[4]

[1]Finnish Research and Engineering, Jaalaranta 9 B 42, 00180 Helsinki, Finland
[2]Center for Nanophase Materials Sciences, Oak Ridge National Laboratory, Oak Ridge, Tennessee 37831, USA
[3]Department of Chemistry and 4D LABS, Simon Fraser University, Burnaby, British Columbia V5A 1S6, Canada
[4]Clarendon Laboratory, Department of Physics, University of Oxford, Parks Road, Oxford OX1 3PU, United Kingdom

### *Abstract*

A single crystal of lead-zirconate-titanate (PZT), composition $Pb(Zr_{0.80}Ti_{0.20})O_3$, was studied by polarized-Raman scattering as a function of temperature. Raman spectra reveal that the local structure deviates from the average structure in both ferroelectric and paraelectric phases. We show that the crystal possesses several, inequivalent complex domain boundaries which show no sign of instability even 200 K above the ferroelectric-to-paraelectric phase transition temperature $T_C$. Two types of boundaries are addressed. The first boundary was formed between ferroelectric domains below $T_C$. This boundary remained stable up to the highest measurement temperatures, and stabilized the domains so that they had the same orientation after repeated heating and cooling cycles. These domains transformed normally to the cubic paraelectric phase. Another type of boundary was formed at 673 K and exhibited no signs of instability up to 923 K. The boundary formation was reversible: it formed and vanished between 573 and 673 K during heating and cooling, respectively. A model in which the crystal is divided into thin slices with different Zr/Ti ratios is proposed. The physical mechanism behind the thermal-stress-induced structural changes is related to the different thermal expansion of the slices, which forces the domain to grow similarly after each heating and cooling cycle. The results are interesting for non-volatile memory development, as it implies that the original ferroelectric state can be restored after the material has been transformed to the paraelectric phase. It also suggests that a low-symmetry structure, stable up to high-temperatures, can be prepared through controlled deposition of layers with desired compositions.

### *Introduction*

Deviations from average crystal symmetry are frequently considered as defects. Certain point defects, such as vacancies are energetically favorable and occur in all crystals [1]. Ferroelectric and magnetic materials possess technologically important two-dimensional defects, that is domain walls. Understanding the formation and motion of domain walls is still a challenging and active research and engineering challenge. The response of ferroelectric domain walls to external stimuli, such as electric field or stress, dictates how quickly and how often the polarization can be switched between different states. The domains form once symmetry lowers from a real (or less seldom, hypothetical) high-symmetry phase at a phase transition. Within Landau's phase transition theory, each domain corresponds to a minimum of the thermodynamic potential. Assuming that the ferroelectric and high-symmetry phase are related by a subgroup-group relationship the symmetry elements lost at a phase transition interconvert the structures of different domains [2]. Ideally this results in domain boundaries which also possess symmetry elements, such as mirror planes. Such domain (twin) boundaries fulfill the mechanical compatibility requirement. Another requirement in ferroelectric materials is that the boundary should be charge neutral [2,3]. It is frequently the point defects at high-symmetry phase that trigger the forthcoming domain pattern formation and serve as domain nucleation and pinning centers in the ferroelectric phase. The interplay between the defects, domain growth and the electric field of the atomic force microscope has resulted in a large research and engineering field of its own, an example being piezoresponse force microscopy [4]. The domain boundary can be quite complex and not really a two-dimensional surface. The crystal structure of solid solutions is often very complex, as is demonstrated by the presence of mixed phases in lead-zirconate-titanate $Pb(Zr_xTi_{1-x})O_3$ (PZT) for $x > 0.50$ [5,6,7,8].

Here we show that thermal-stress-induced boundaries can be very stable and remain up to several hundred degrees above the phase transition temperature. The present study is dedicated to rhombohedral, $x = 0.80$, PZT single crystal. We apply a very sensitive and fast high-resolution spectroscopic technique to address the structural properties, with the focus being on the short-range order in the rhombohedral and cubic phases.

### *Experimental*

***Sample preparation.*** Single crystals of $Pb(Zr_{0.80}Ti_{0.20})O_3$ were grown by a top-seeded solution growth technique using a mixture of PbO and $B_2O_3$ as flux at the Simon Fraser University. A crystal platelet of pseudocubic (001) orientation with dimensions 2.12 mm x 1.56 mm x 0.60 mm was prepared with the largest (001) faces mirror-polished. The two long edges were parallel to the pseudocubic *a* and *b* axes and the short edge was parallel to the *c* axis.

***Raman experiments.*** Raman measurements were performed using a Jobin-Yvon T64000 spectrometer consisting of a double monochromator coupled to a third monochromator stage with 1800 grooves per millimeter grating (operating in a double subtractive mode). The acquisition time was adjusted to have a sufficient signal-to-noise ratio. A liquid nitrogen-cooled charge-coupled device detector was used to count photons. All measurements were carried out under a microscope in the backscattering configuration. The Raman spectra were excited using a continuous wave solid-state laser (wavelength 532 nm). The laser beam power on the sample surface was 2 mW, and the diameter of the laser beam spot was approximately 2 μm. The laser light was focused on a sample surface by a long focal length objective (LM Plan F1100x, N/A = 0.8). A Linkam stage (model TS-1500) and a temperature control unit (model TMS-94) were used for the room and high-temperature (between 300 and 800 K) measurements. The spectra were not divided by the Bose-Einstein thermal factor as there is no *a priori* knowledge justifying the assumption that the signal was solely due to first-order Raman scattering. This is evidently crucial for solid solutions at high-temperatures. A silicon peak was used to correct for peak positions. The correction was typically less than 0.5 $cm^{-1}$. The Raman spectra were analyzed using Jandel Scientific PeakFit 4.0 software.

***Results***

Dielectric measurements revealed that the phase transition between the high-temperature $Pm\bar{3}m$ and $R3m$ phases occurred at 573 K, and the transition between the $R3m$ and $R3c$ phases occurred at 403 K. The transition temperatures were in excellent agreement with earlier reports [9].

For backscattering measurements the crystal was oriented so that the laser beam was parallel to the pseudocubic *c*-axis and the polarization of the laser light was parallel to the pseudocubic [110] direction. This configuration allows the detection of the *E*-symmetry modes alone, which was the prime reason for choosing this geometry. To gain a better understanding of the boundaries described below, experiments were also conducted with the laser light polarization direction parallel to the pseudocubic [100] direction. As is discussed below, the evolution of the spectra as a function of temperature was mainly consistent with the known phase transition sequence $R3c \rightarrow R3m \rightarrow Pm\bar{3}m$. However, two sharp Raman peaks and a broad band between 180 and 320 $cm^{-1}$, still present at 773 K, were observed, 200 K above the ferroelectric to paraelectric phase transition. These features were accompanied by the formation of a boundary region, shown in Fig. 1 (a). To interpret the Raman spectra, four types of domains were considered, indicated by pseudocubic axes $\boldsymbol{a}_{R1}$+, $\boldsymbol{b}_{R1}$+, $\boldsymbol{a}_{R1}$,- and $\boldsymbol{b}_{R1}$- (left-hand side) and $\boldsymbol{a}_{R2}$+, $\boldsymbol{b}_{R2}$+, $\boldsymbol{a}_{R2}$,- and $\boldsymbol{b}_{R2}$-, where + and - indicate that the pseudocubic $\boldsymbol{c}_R$ axis points upwards or downwards, respectively. The Raman scattering tensors corresponding to each domain were expressed in terms of the coordinate axes $x'$, $y'$ and $z'$, where $x'$ corresponds to the light polarization direction and $z'$ pointing upwards being parallel to the laser beam propagation direction, see Appendix I. In the text we refer to the four domain candidates as 1+, 1-, 2+ and 2-. Both peaks were observed with crossed polarizers, whereas only the highest-frequency peak was observed with a parallel polarizer. The stripes appeared in the vicinity of the domain boundary, as was confirmed by measuring spectra with parallel and crossed polarizers at 303 K from the left and right-hand side of the boundary region, shown in Fig. 2.

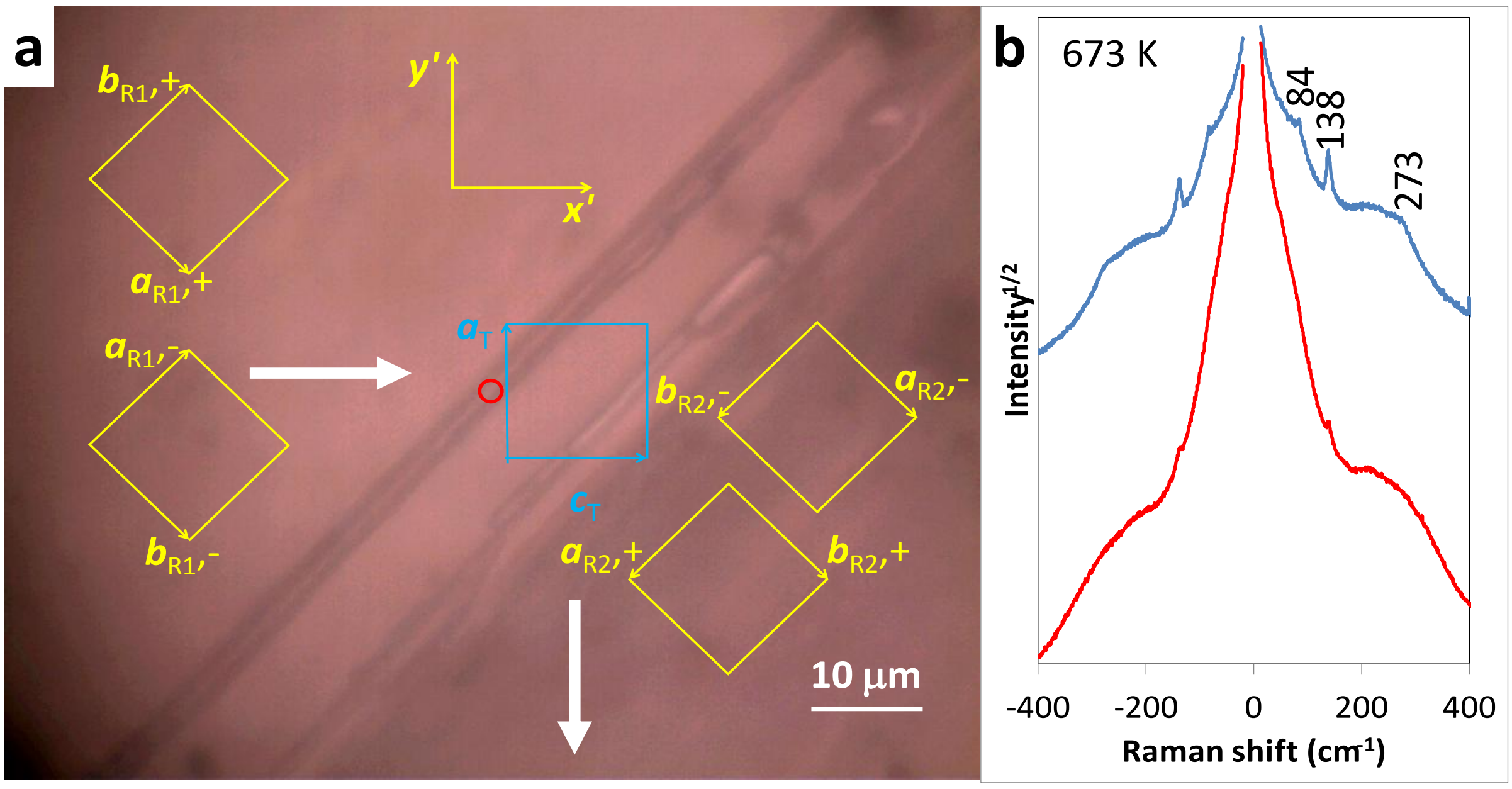

Fig. 1. (a) Pseudocubic (001) plane at 673 K. Diagonally propagating stripes became visible at around 470 K and were assigned to a domain (twin) boundary. The white arrows indicate the projection of the polarization vectors when the crystal is cooled below 573 K. The spot at which the laser beam was focused is shown by a red circle. (b) Stokes and anti-Stokes Raman spectra measured at 673 K from the boundary area. The blue line spectra were measured with crossed polarizers, and the red line spectra by parallel polarizers. Sharp peaks characteristic of the boundary region are seen at 84 and 138 $cm^{-1}$. The square root of the intensity is shown in order to reveal the weak features.

The intensity ratios of the Raman peaks of the two domains are different. The symmetry analysis for the four domain candidates is given in Appendix I. Thus, we refer to this boundary as a domain boundary, though it actually consists of several nearly parallel slices. The unexpected features shown in Fig. 1(b) are discussed in the context of the cubic phase.

Though the data do not allow us to conclude the exact polarization of the domains, essential information can be extracted. First, we assume that the boundary is not charged and separates areas whose polarization vector components on the (001) plane differ by 90°, implying that one domain is of the type 1±, the other is of the type 2± (other cases correspond to sign changes of the polarization vector). Reference to Table A3 suggests that the left-hand side domain is either 1+ or 1- and the right-hand side domain is 2+.

Next we discuss the features measured from the rhombohedral phase, and the phase transition to the cubic phase.

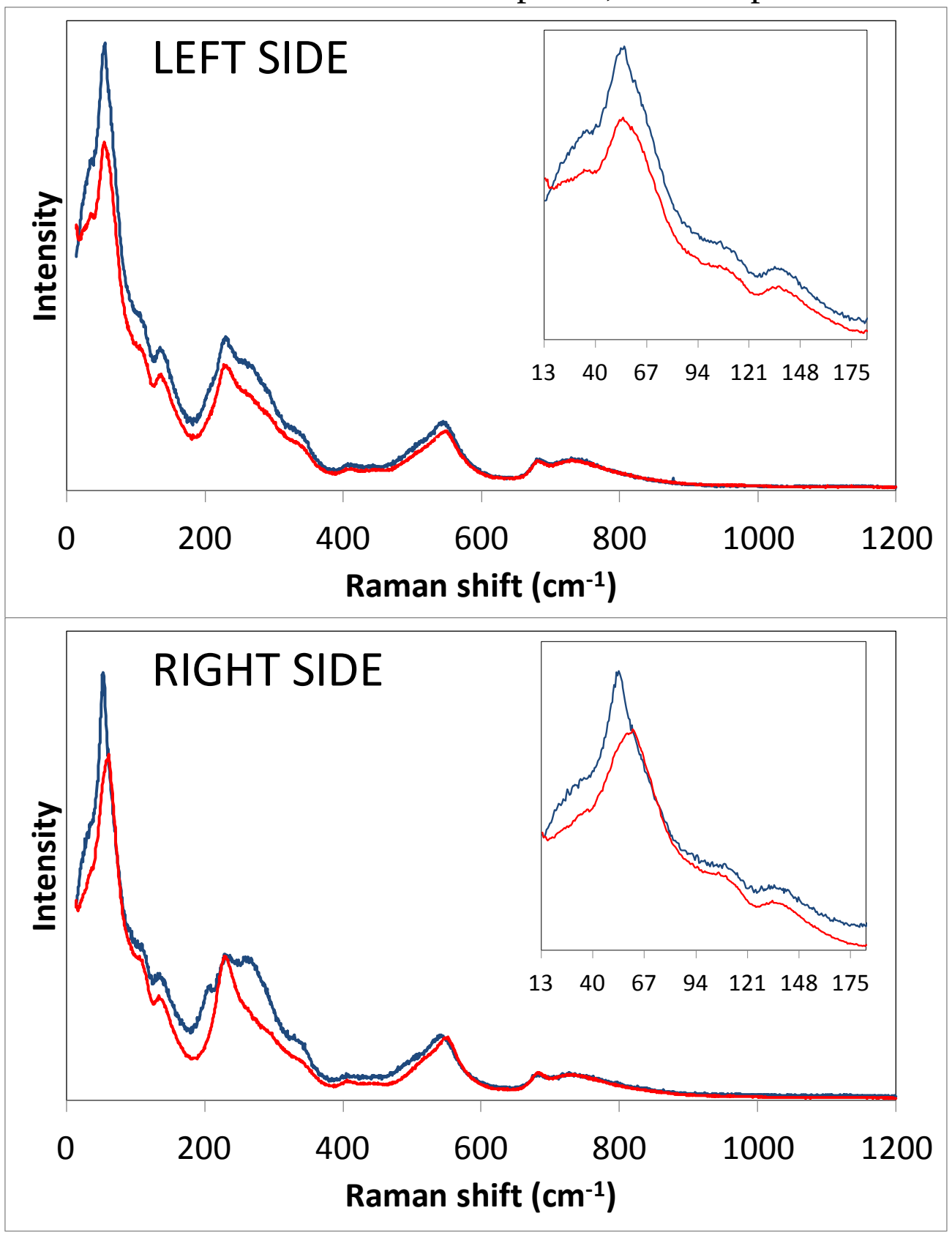

Fig. 2. Raman spectra measured from the left- and right-hand side of the domain boundary at 303 K. The blue line spectra were measured with crossed polarizers, and the red line spectra by parallel polarizers. By symmetry, the blue line spectra show only *E*-symmetry modes whereas the red line spectra show both *E*- and $A_1$ symmetry modes (see Appendix I). In solid solutions this assignment is not strictly true because of the large number of crystalline defects. The left-hand side domain is rather insensitive to the polarizer position, whereas the right-hand side domain shows more evidence of polarization dependence. The intensities were scaled so that the band between 650 and 900 $cm^{-1}$ had equal intensities.

***R*3*c* and *R*3*m* phases.** Curve fitting of the room-temperature spectra revealed that the number of Raman-active modes, 18 in Fig. 3, was larger than expected by the symmetry. Two of the modes, weak peaks at 408 and 438 $cm^{-1}$, are assigned to longitudinal modes. Also the band between 650 and 900 $cm^{-1}$ presumably contains the longitudinal highest frequency $A_1$ and *E*-symmetry modes, see Appendix I, section D. According to symmetry analysis, the Raman-active modes transform as $4A_1$+9*E*. In the measurement geometry applied to obtain the spectra in Fig. 3 only *E*-symmetry modes are allowed: with crossed polarizers $A_1$-symmetry modes are not expected to occur, though this selection rule is not strictly obeyed in PZT solid solutions, as discussed in ref. 10. The larger than predicted number of modes suggests that the origin of the peak splitting is due to the splitting of the *E*-symmetry modes.

The main difference between crossed and parallel polarization geometries is observed on the right-hand side domain: the peak at around 60 $cm^{-1}$ consists of two peaks, one of which nearly disappears with crossed polarizer measurement. Figs. 3 (a), (c) and (e) show the room-temperature spectra (*R*3*c* phase) and curve fits measured from the right-hand side domain, and Figs. 3(b), (d) and (f ) show the corresponding data at 433 K (*R*3*m* phase). Curve fitting was conducted also for the room-temperature spectra collected with parallel polarizer geometry. The band between 180 and 380 $cm^{-1}$ became simpler with parallel polarizer geometry, the mode at 205 $cm^{-1}$ vanished and the modes approximately at 260, 300 and 340 $cm^{-1}$ had significantly lower intensity.

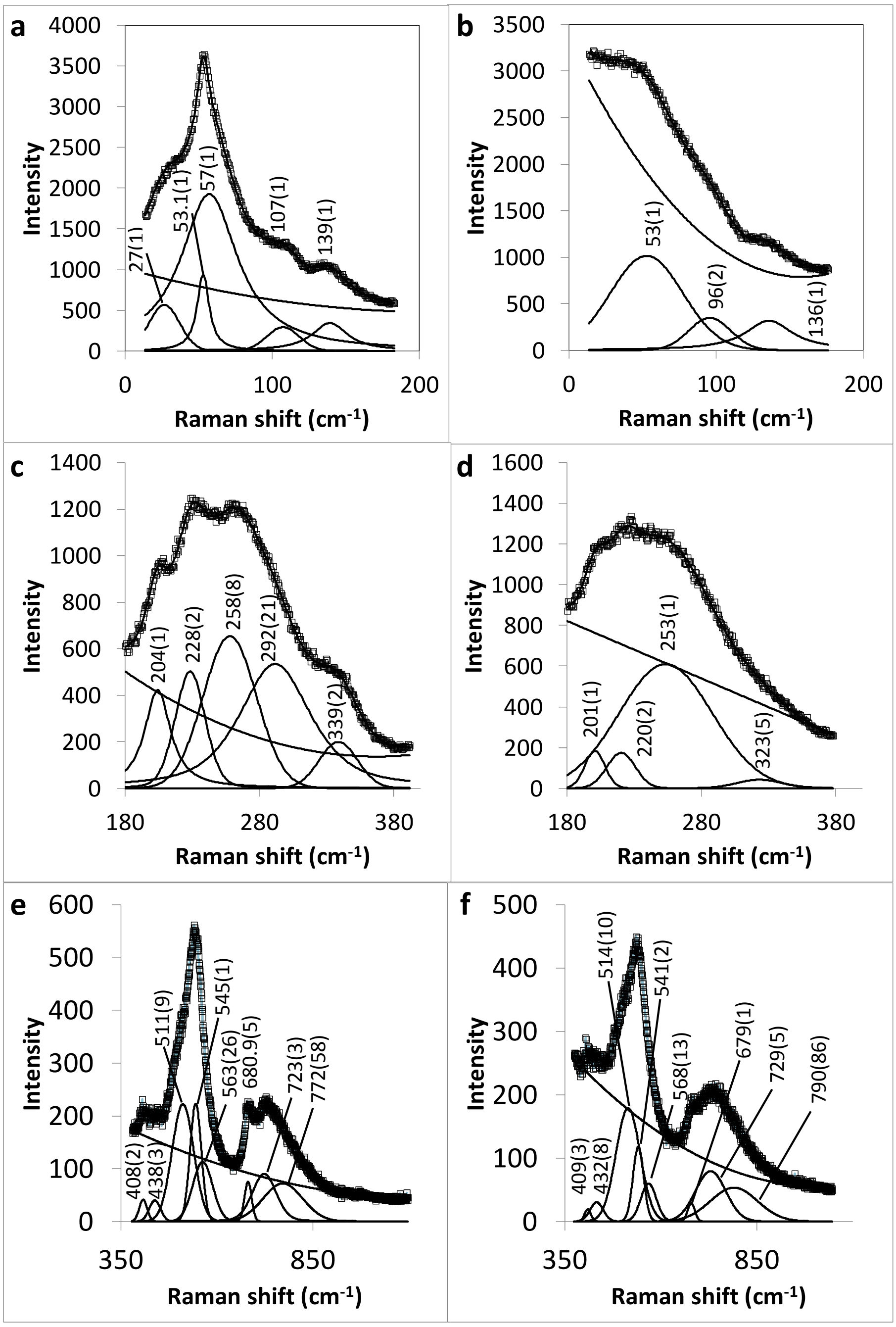

Fig. 3. Left column shows the Raman spectra measured at 303 K with crossed polarizers from the right-hand side domain, phase *R*3*c*. Right column shows the spectra measured at 433 K, phase *R*3*m*.

Figs. 4(b) and (c) show that the spectra seem to become continuously simpler with increasing temperature. However, this impression is mainly due to the broadening of the peaks, except that for the lowest-frequency soft modes the true number of the peaks is almost the same at 433 and 303 K, as Fig. 3 reveals. In the intermediate Raman shift range (Figs. (c) and (d)) the peaks at 258 and 292 at room temperature could equally well be fitted by a single peak at 433 K, as was done in Fig. 3(d). Thus, the number of peaks is still much higher in the *R*3*m* phase than expected by the symmetry. For an ideal *R*3*m* symmetry the Raman-active modes transform as $A_1$+4*E*, so that only four modes are expected for this geometry. This is consistent with the idea suggested earlier [11,12,13,14] that Pb-displacements have a crucial role for local structural distortion. A recent review of the role of Pb-displacements in piezoelectric perovskites is given in ref. 15. For Ti-rich compositions Pb displacements are essentially distributed along the ⟨11*z*⟩ directions, whereas they are expected to be along one of the ⟨111⟩ directions (the hexagonal *c*-axis)in the ideal rhombohedral PZT. However, neutron and transmission electron microscopy studies have indicated that Pb is displaced in directions perpendicular to the hexagonal *c*-axis [11,12]. The displacements are spatially distributed and do depend on the local environment, mainly on the occupation of the eight nearest *B*-cations. A recent Pair Distribution Function and Rietveld refinement study has shown that Zr-rich Pb($Zr_xTi_{1-x}$)$O_3$ powders possess mixed phases, described by *R*3*c*/*R*3*m*+*Cm*($M_B$) model for $0.65 < x < 0.92$ and *R*3*c*/*R*3*m*+*Cm*($M_A$) model for $0.52 < x < 0.65$, where $M_A$ and $M_B$ refer to two polarization direction variants of the *Cm* phase [16]. For $x = 0.80$ the Pb displacement direction of the majority rhombohedral phase with respect to the rhombohedral [111] direction is relatively isotropic whereas the Zr displacement directions are highly disordered [16]. This is consistent with the large number of very broad Raman peaks that are often obtained, notably seen as a splitting of the *E*-symmetry modes that are present also in the higher-symmetry *R*3*m* phase.

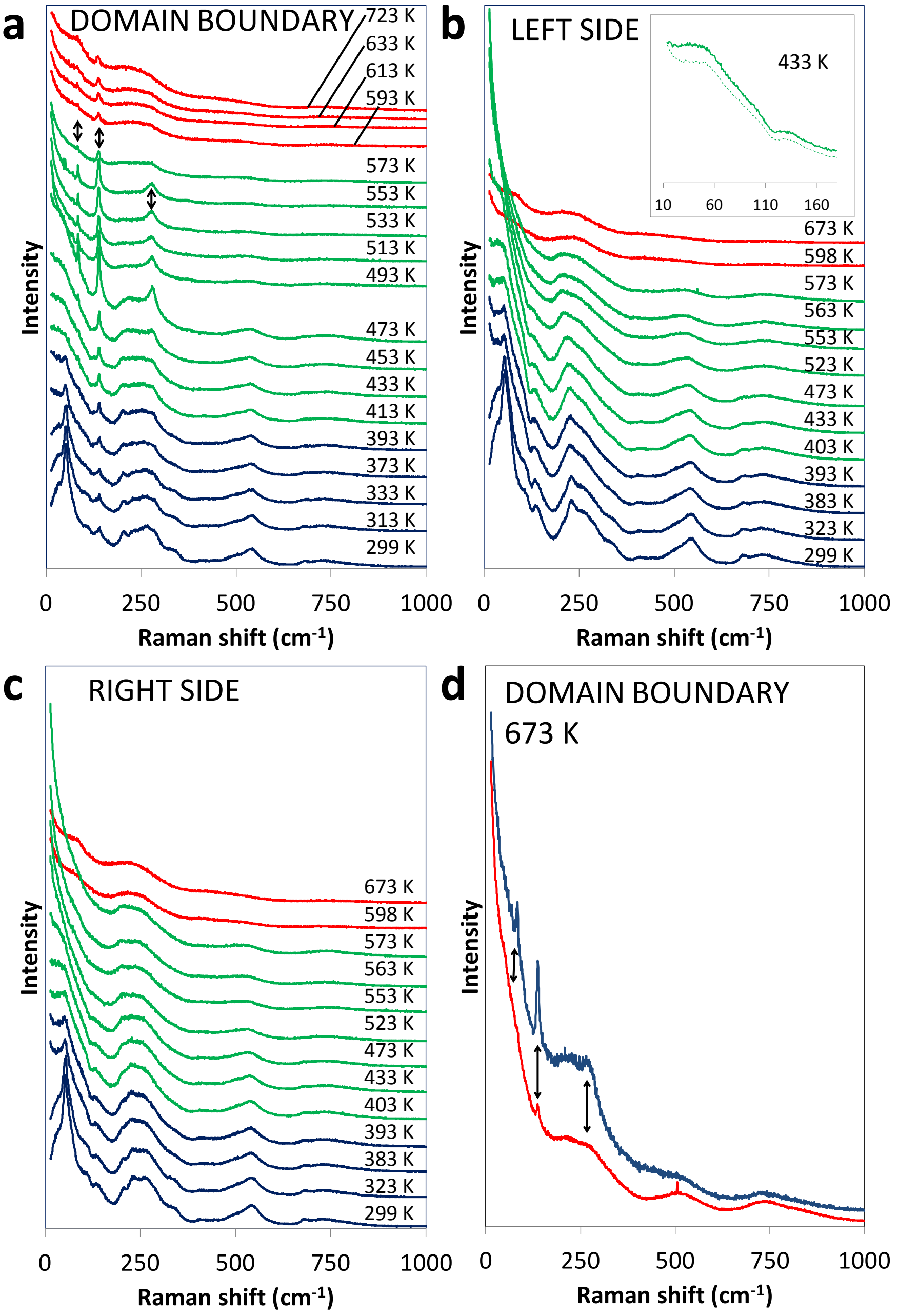

Fig. 4. (a) Temperature dependence of the Raman spectra collected in the boundary area. The spectra are a mixture of the signal coming from the domain wall and domain area. Arrows show features characteristic of the domain wall. Note that these features are seen in the high-temperature phase, which nominally has $Pm\bar{3}m$ symmetry with no Raman-active modes expected. The spectra collected on the left- and right-hand side domains after repeated heating and cooling are given in panels (a) and (b), respectively. Spectra corresponding to the $R3c$, $R3m$ and $Pm\bar{3}m$ phases are plotted by blue, green and red lines, respectively. The spectra shown in panels (a)-(c) were measured with crossed polarizers, except for the inset in panel (b) which shows spectra collected with crossed (continuous line) and parallel polarizer (dashed line). Panel (d) shows the spectra collected with crossed polarizers (blue line) and parallel polarizers (red line) at 673 K in the same measurement cycle as the spectra shown in panels (b) and (c). All domain-wall signatures were weaker in spectra measured with parallel polarizers.

Even though the symmetry increases at the $R3c \rightarrow R3m$ transition, the short-range structure significantly deviates from long-range order as is seen from the curve fit shown in Fig. 3. From the symmetry viewpoint, the transition $R3m \rightarrow R3c$ corresponds to an instability of the Brillouin zone corner point [17], resulting in a doubling of the primitive cell [18]. The most evident change occurring in the transition is the gradual disappearance of some of the low-frequency modes, compare Fig. 3 (a) and (b). The computational data provided in section D of Appendix I give a qualitative explanation for this: in the $R3m$ phase only two modes (weak LO and stronger TO $E$-symmetry mode) are expected below 200 $cm^{-1}$, whereas 7 modes (weak LO and stronger TO $E$-symmetry modes) are expected in the $R3c$ spectrum collected with crossed polarizers. One TO and LO $A_1$ mode pair is expected by symmetry (see Fig. A11(a) and Table A4) which should appear in the spectrum collected with parallel polarization. As the inset in Fig. 4(b) shows, the spectra measured with crossed and parallel polarizers are similar suggesting that the contribution of $A_1$-symmetry modes is not significant. This could be due to the high temperature which softens and broadens the lowest-frequency modes. The computational results given in Appendix I correspond to the ground state. The transition is seen to be continuous, consistent with the earlier neutron diffraction work, in which the continuous transition was interpreted in term of a weak coupling between oxygen octahedral tilting and octahedral strain [12]. In contrast, the tilt in $Pb(Zr_{0.90}Ti_{0.10})O_3$, which is close to the Zr-rich end of the rhombohedral range, goes discontinuously to zero at the $R3c \rightarrow R3m$ transition [19].
The domain formation in PZT has been suggested to depend on the spatial distribution of Zr and Ti ions which in turn dictates the Pb-displacements [10]. This generates small ordered regions within domains possessing long-range order. In tetragonal PZT the main component of ferroelectric polarization is along the $c$-axis, which is the long-range feature. However, within the domain the local Pb-displacements are incoherently distributed along the $\langle 11z \rangle$ directions, corresponding to local monoclinic symmetry, since the ferroelectric polarization favors all four sites equally. The physical origin of off-center Pb-displacements is due to the 6$s$ lone-pair electrons which tend to enhance the existing distortion. In PZT the distortion is proportional to the spontaneous polarization. It is a commonly adopted view that the Pb-cations easily respond to external stimuli, electric field or stress, which makes PZT a good piezoelectric and ferroelastic material.

**$Pm\bar{3}m$ phase.** The transition to the cubic phase was revealed by the softening of the lowest-frequency modes, which nearly vanished at 523 K below the central peak, Figs. 4(b) and (c). Also the spectra measured from the left- and right-side of the boundary became indistinguishable in the cubic phase, see the red spectra in Figs. 4(b) and (c). Because the Raman shifts of the soft modes become very small and are masked by the strong Rayleigh scattering at high temperature, it is difficult to follow their temperature dependence. Correspondingly, other phase transition dependent properties were sought. The central peak provided the most evident sign of the $R3m \rightarrow Pm\bar{3}m$ transition: the intensity increased rather rapidly when the transition was approached and decreased quickly after the transition point was exceeded, see Fig. 5(d). This is consistent with the notion that the first-order phonon contribution to the optical dielectric constant vanishes in the centrosymmetric phase [3,20]. Recalling that the response of the crystal to the incoming electric field of the light can be expressed in terms of an optical dielectric constant ($\varepsilon$) modulated by phonons, following ref. 3 one can write down a series expansion of the dielectric constant with respect to polarization, $\varepsilon = \varepsilon_0 + \left(\frac{\partial \varepsilon}{\partial P}\right) P + \frac{1}{2}\left(\frac{\partial^2 \varepsilon}{\partial P^2}\right) P^2 + \cdots$ . This expansion allows the consideration of the high-temperature phase potentially possessing local polarization regions. The first term is zero in the centrosymmetric phase and thus the second- and higher-order terms are responsible for the possible increase in $\varepsilon$. If the second- and higher-order terms are small, the drop in the central peak intensity is abrupt. In the case of an order-disorder phase transition there still is a multiphonon contribution to the optical dielectric constant as there are local, spatially disordered dipole moments. We note that Pb-ions are highly polarizable and are also known to be disordered at high-temperature. Though there are probably not sufficient numbers of data points for the cubic phase, it is worth considering the possibility that the slight increase in the central peak intensity with increasing temperature is due to the order-disorder character of Pb-ions or oxygen. A recent high-resolution neutron powder diffraction study showed that the routinely used $Pm\bar{3}m$ symmetry does not give a satisfactory fit for the high-temperature phase of Zr-rich PZT [21]. The origin of the features that are not explained by the ideal $Pm\bar{3}m$ symmetry are related to remnant, either dynamic or static, tilting of oxygen octahedra. Note that our measurements were conducted with crossed polarizers, which means that the susceptibility tensor describing the Rayleigh scattering has a non-diagonal component. The signature was reversible and the maximum intensity was at around 563 K, consistently with the dielectric measurements and phase diagram [9]. Interestingly, the central peak area measured from the boundary region showed only a weak anomaly, possibly resulting from scattering from neighboring domains. This further supports the idea that the boundary region exhibits no phase transition in the temperature range studied.
A peculiar feature was that the boundary was not visible at room-temperature, but became visible at around 450 K and was accompanied by a strengthening of the peaks at around 84, 138 and 273 $cm^{-1}$, as shown in Fig. 3a. We interpret that this as coming from thermal stress developing in the domain wall area. It is worth noting that the areas around the boundary behave normally and exhibit no peaks in the paraelectric phase region. The domain wall itself seems not to be thin; rather it is a volume consisting of several nearly parallel planes, as Fig. 1 shows. To extract further information curve fitting of the three domain wall peaks were carried out. Fig. 5 shows the peak positions as a function of temperature. Three data sets, collected during different heating cycles, were analyzed. Repeated heating and cooling measurement cycles were applied to test the stress-induced interface-phase model. The results show that no anomaly was observed at the phase transition temperatures at 403 and 573 K (see Fig. 5 (b)). Instead, the peak frequencies decreased approximately linearly with increasing temperature. The overall behavior remained the same, indicating that the structural changes occurring in the domain wall were reversible, as is evident from Fig. 5. The domain orientations were preserved after repeated heating and cooling cycles, even though the highest temperatures were 773 K, 200 K above the ferroelectric-paraelectric phase transition temperature.

In principle, the interface between the two domains may be a mirror plane, connecting the two crystalline parts without a mismatch. In the present case a volume with a different structure was formed parallel to the pseudocubic (100) plane, which in turn separated two differently oriented crystal sections. The phase has spectral features similar to the one found in PZT oxides and matches with the peak positions reported for $PbZrO_3$ [22]: 82, 137 and 279 cm$^{-1}$. The polarization dependence implies that the peaks are not totally symmetric $A_1$-modes as they nearly disappear with parallel polarizer geometry. In addition to the 16$A_{1g}$ symmetry modes, $PbZrO_3$ has 16$B_{1g}$ (*xy*-geometry), 14$B_{2g}$ (*xz*-geometry) and 14$B_{3g}$ (*yz*-geometry) modes [23], so that the modes could either posses $B_{1g}$ or $B_{2g}$ symmetry [24]. The strong enhancement of the intensities of the peaks at around 84, 138 and 273 cm$^{-1}$ suggests that the stress increases with increasing temperature and decreases the symmetry of the domain boundary region. Thus, a structural model in which a slice of $PbZrO_3$ phase is embedded between the domains is not correct, as there are no signs of the $PbZrO_3$ phase at room temperature. A further puzzling point is that only three modes were observed at elevated temperatures, whereas several more modes should be visible. The third point worth to note is that in $PbZrO_3$ the transition to the cubic phase should take place at 503 K. Based on this we conclude that the volume between the two domains is a perovskite structure under severe stress. To test this model a set of experiments was conducted.

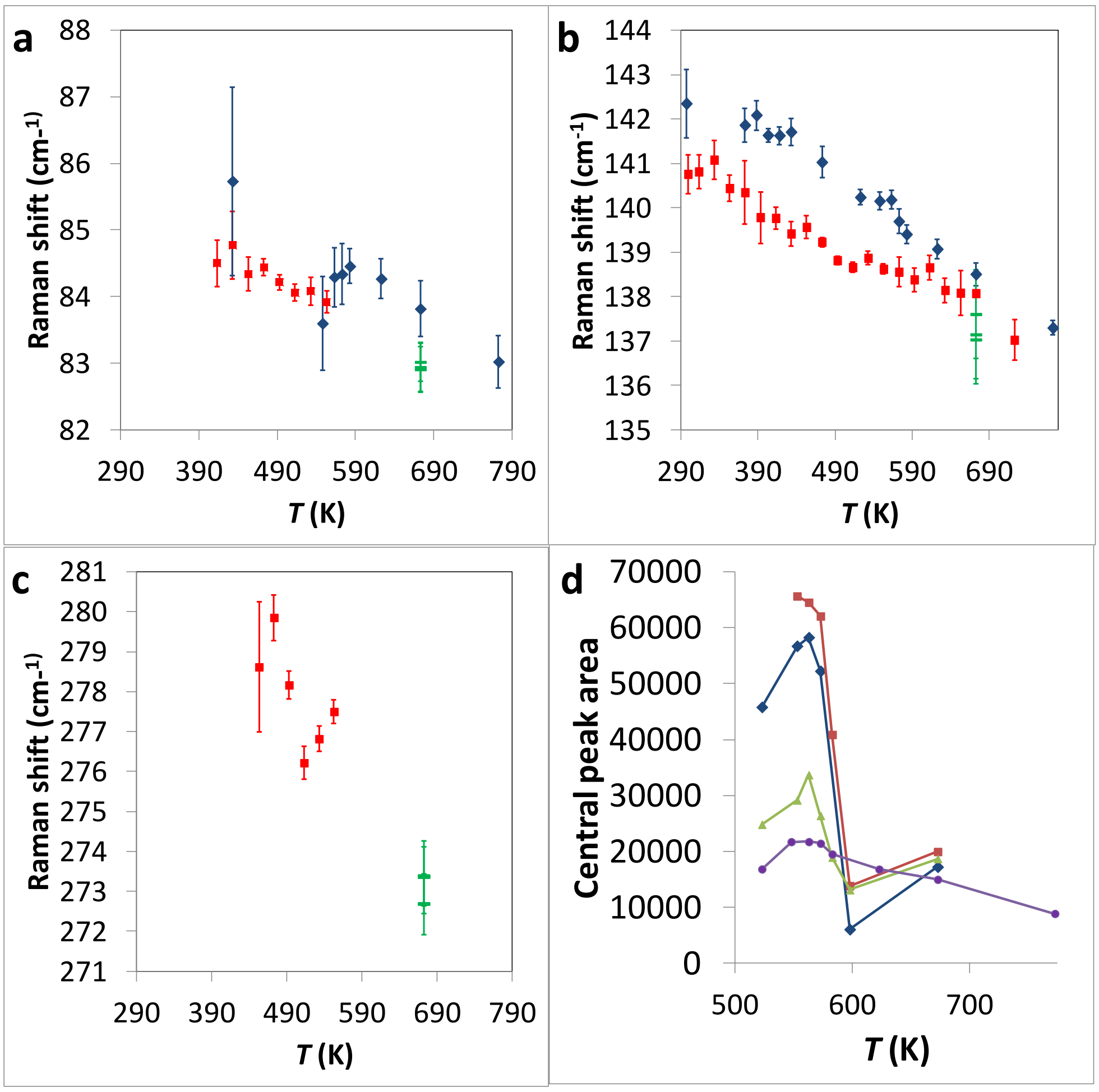

Fig. 5. Raman shift of the domain wall boundary modes as a function of temperature. 95% confidence limits are also indicated, reflecting the quality of the fit. In panels (a), (b) and (c) different colors indicate data collected on different days: red data was measured first and blue and green data one and two days later. The small difference in peak positions is assigned to a slightly different spatial location and changes in the thermal stress state of the boundary. Panel (d) shows the area under the Rayleigh peak, measured with crossed polarizers. Blue and red data points were measured from the left-hand side domain during cooling and heating, respectively, and green data points were measured from the right-hand side domain during cooling. Purple data points (filled spheres) show the area measured from the boundary

region during heating.

Figure 6 shows the Raman spectra collected on a stripe similar to the one shown in Fig. 1. The incoming laser beam polarization is perpendicular to the stripe line. As in Fig. 1, the crossed polarizer measurements show sharp features, whereas with the polarization component parallel to the incoming laser beam polarization there is nearly zero intensity. Measurement on a rupture nearly parallel to the stripes shows no Raman signal at 673 K, Fig. 6 (d). This is consistent with the idea that thermal stress causes the symmetry-lowering resulting in the extra sharp peaks.
A series of measurements up to 923 K were conducted in the vicinity of the ruptured area, Fig. 7. The rupture shown in Fig. 7 (a) (right-hand side) is parallel to the features shown in Figs. 1 and 6 and was chosen as in the vicinity of the rupture the crystal can expand more freely than inside an undamaged domain. As in crystal portions shown in Fig. 1 (a) and Fig. 6 (a), faint stripes appeared at around 673 K. The angle between the stripes shown in Fig. 1(a) and 6(a) is approximately 45° with respect to the stripes shown in Fig. 7(a). The absence of the stripes parallel to the rupture is assigned to the stress relief via rupture. The rupture is not able to relieve stress assigned to the stripes that are not parallel to the rupture line. One of the lines is emphasized by the dotted line in Fig. 7 (a). The formation of stripes was again reversible; Fig. 7 (d) shows that the features disappear during cooling. As in the case of modes shown in Fig. 1 the temperature dependence of the mode frequencies is linear, though now the higher-frequency mode at around 80 $cm^{-1}$ increases with increasing temperature, whereas opposite behavior is observed in the mode at around 50 $cm^{-1}$. The peak widths slightly increased with increasing temperature (Fig. 7 (b)), though they are sharp when compared with the bands between 180 and 400 $cm^{-1}$, 400 and 610 $cm^{-1}$ and 610 and 900 $cm^{-1}$: compare with the high-temperature spectra shown in Fig. 4(b) and (c) The changes in these regions are also visible, though less evident than the low-frequency modes. This suggests that thermal stress-induced symmetry-lowering is the origin of the sharp, high-temperature features.

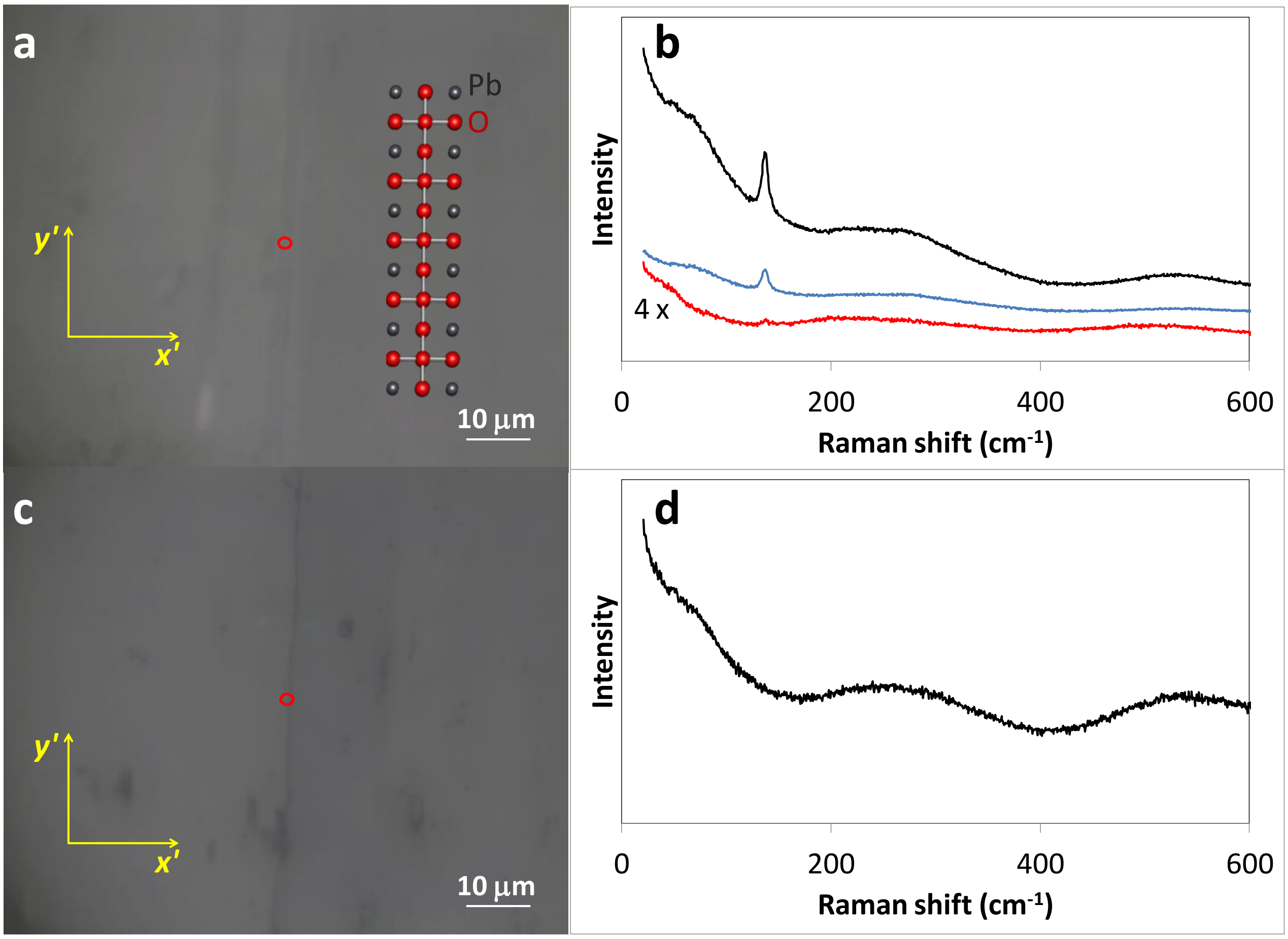

Fig. 6. Panels (a) and (b) show the stripes and the corresponding Raman spectra at 673 K. The blue line spectra were measured with crossed polarizers and the red line spectra by parallel polarizers (intensity multiplied by a factor 4). The structure in (a) is the arrangements of atoms in the non-deformed crystal. The spectra plotted in black in panels (b) and (d) were measured without polarizers. The spot at which the laser beam was focused is shown by the red circle in panels (a) and (c). Panel (d) shows the Raman spectrum collected on the rupture line, panel (c), at 673 K. The spectrum indicates that after the stress is relieved the sharp features also disappear. *x'* indicates the light polarization direction and *z'* points upwards, being parallel to the laser beam propagation direction.

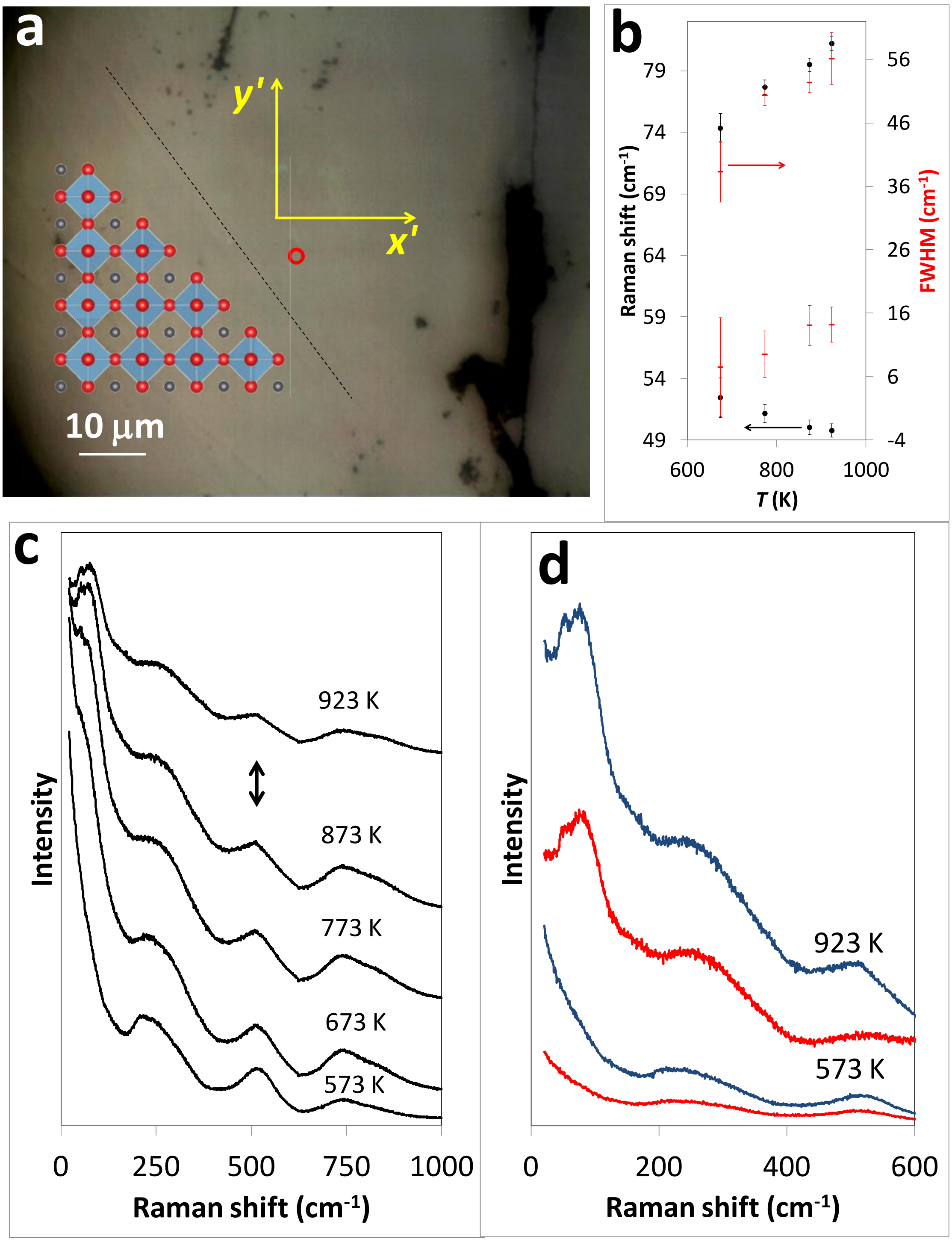

Fig. 7. Panel (a) shows the stripes formed in the vicinity of the rupture area at 873 K. At around 673 K a new set of stripe formations, indicated by the dotted line in panel (a) was observed, accompanied by Raman peaks shown in panels (c) and (d). The crystal structure of the non-deformed crystal is drawn in (a). The two-headed arrow in panel (c) shows that also the some hard modes are sharpened. Panel (d) indicates that the structural deformation is reversible, as the spectra measured during cooling (spectra measured at 573 K are shown) has no signature of first-order Raman scattering. The blue line spectra were measured with crossed polarizers, and the red line spectra by parallel polarizers. The spot at which the laser beam was focused is shown by the red circle in panel (a). Panel (b) shows the result of curve fitting, which indicates that the frequency of the sharper and lower-frequency mode decreases with increasing temperature, whereas the frequencies of the higher frequency modes increase with increasing temperature. The red data points give the full-width-at-half-maximum values of the peaks. *x'* indicates the light polarization direction and *z'* points upwards, being parallel to the laser beam propagation direction.

**Structural origin of the strain-induced symmetry lowering.** The behavior of the additional peaks can be understood by thermal stress-induced symmetry-lowering. At first sight it might seem surprising that the symmetry-lowering becomes evident well above the $R3m \rightarrow Pm\bar{3}m$ phase transition. The fact that the peaks appear at the same (difference being a few $cm^{-1}$ and always less than 10 $cm^{-1}$) Raman shift values as the first-order Raman modes of the high-temperature $R3m$ phase strongly suggests that the high-temperature peaks correspond to the distorted perovskite phase. A model based on the statistical distribution of Zr and Ti ions is outlined in Appendix II. The key point is that the crystal can be divided into parallel thin slices. With a random distribution of Zr and Ti ions there is a variation between the average compositions of slices which in turn means that each slice has its own specific physical properties, such as thermal expansion tensor. This suggests that there is a limited temperature range in which the slices are smoothly joined together. We also note that it is not only the slice composition but also the boundary conditions (e.g., the mechanical clamping or absence of mechanical clamping in one direction which can happen in the vicinity of a rupture). Furthermore, different crystal directions yield more easily than the others so that the choice of slice orientation depends on the boundary conditions (e.g., clamping) and materials properties. Thus, to model the behavior observed in Figs. 1, 4 and 6(a) and (b) the crystal should be divided into slices parallel to the cubic (100) plane, whereas it should be divided into slices parallel to the (110) planes in order to model the features observed in Fig. 7, as sketched in Fig. A12 in Appendix II.

We make a brief note on the case of powders and potential applications. It is possible that powders do not exhibit similar behavior as the grain size is significantly smaller than the average twin size in the present crystal. Thus, there are numerous paths for a stress relief in powders. However, in thin films controlled composition gradient-engineered deposition techniques allow one to design structures stabilized by strain in the desired temperature range, including temperatures above the known phase transition temperatures.

### *Conclusions*

Raman scattering measurements were conducted on $Pb(Zr_{0.80}Ti_{0.20})O_3$ single crystal. In the ferroelectric phase the number of peaks observed was significantly larger than expected by the space group symmetry and the features remained up to the ferroelectric-paraelectric phase transition point. The rhombohedral $R3c$ and $R3m$ symmetries reflect the average symmetries. Pb-cation shifts and random distribution of Zr and Ti cations result in a multiplicity of different states in a single domain region, which corresponds to the observed Raman peak splitting. The clearest sign of the ferroelectric-paraelectric phase transition was the maximum of the central peak. The nominally cubic paraelectric phase exhibited broad but evident bands showing that the local symmetry deviates from perfect long-range order also at elevated temperatures. Thermal stress-induced structural changes were observed at elevated temperatures and were assigned to the different thermal expansions of the crystal slices possessing different Zr/Ti ratios.

Defects play a significant role for domain formation and a persistent domain boundary was observed that showed no signs of structural transformation during heating well above the ferroelectric-paraelectric phase transition temperature. The boundary locked the two domains so that after repeated heating and cooling cycles to the paraelectric phase the domains returned to the original ferroelectric state, suggesting interesting possibilities for memory cell applications.

## Appendix I: Symmetry analysis

### A. Space group symmetries as a function of temperature

The symmetries of the rhombohedral PZT are given in ref. 25. Above 573 K the average symmetry of $Pb(Zr_{0.80}Ti_{0.20})O_3$ is cubic (space group $Pm\bar{3}m$) and paraelectric. Below 573 K the symmetry is lowered to *R*3*m*, with Pb at 3*a*, *B*-cations at 3*a* and oxygen at 9*b* Wyckoff positions. Below 403 K $Pb(Zr_{0.80}Ti_{0.20})O_3$ adopts *R*3*c* symmetry:the Pb-ions are at Wyckoff position 6*a*, *B*-cations at 6*a* and oxygen at 18*b*. The symmetry-lowering from *R*3*m* to *R*3*c* can be described as $R3m \rightarrow F3c$, which corresponds to a doubling of all rhombohedral axes. The symmetry lowering is caused by $a^{-}a^{-}a^{-}$ tilting of the oxygen octahedra The transition *R*3*m*->*F*3*c* in terms of rhombohedral axes is illustrated in Fig. A8. Since the crystal edges are parallel to the rhombohedral axes of the *R*3*m* phase, it is convenient to use them as a reference frame when describing the experimental geometries. A geometrical approach linking the octahedra tilting and the volumes of the octahedra and cuboctahedra of the *R*3*c* phase is given in ref. 26. Ref. 27 addresses the ionic shifts and the octahedral tilting in the *R*3*c* and $R\bar{3}c$ phases under large hydrostatic pressure (mimicking chemical pressure) in lead titanate.

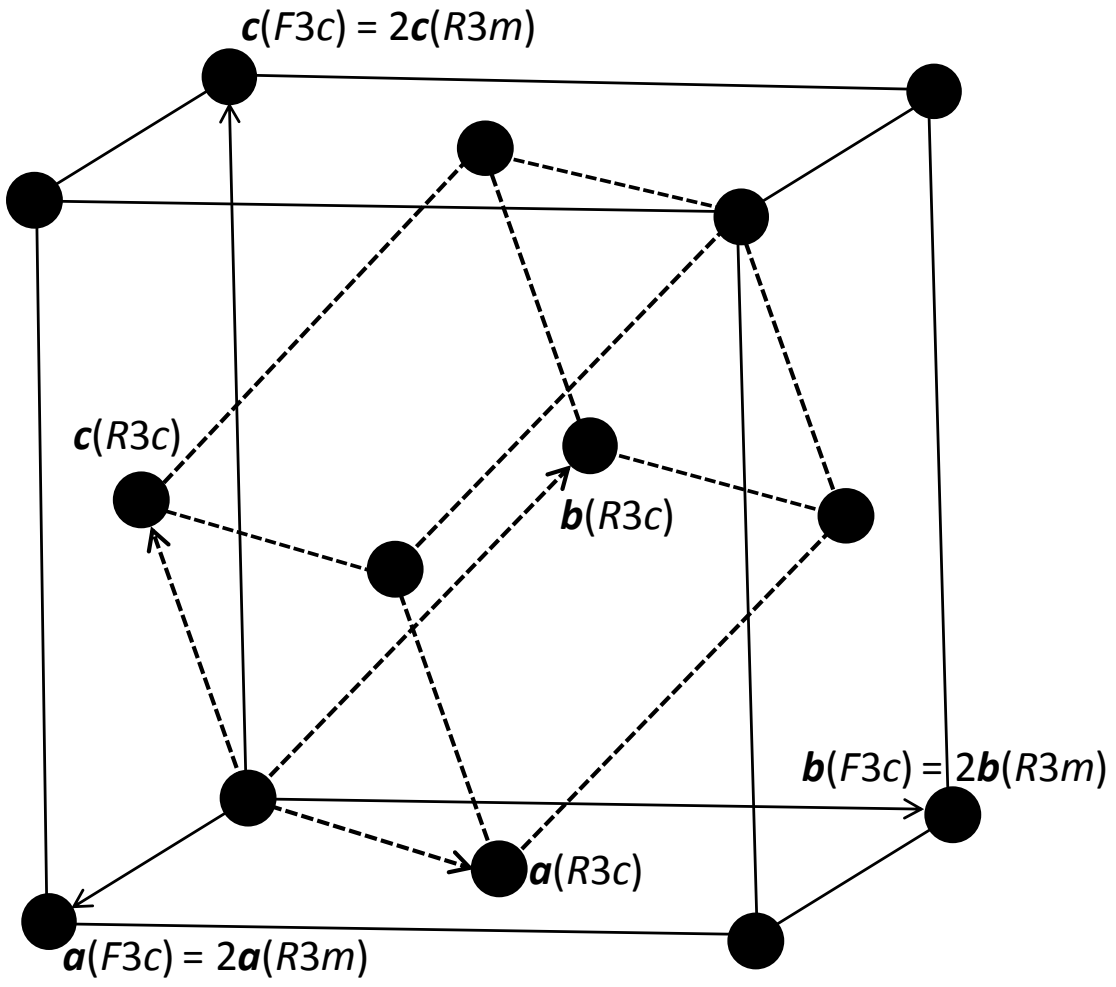

Fig. A8. The relationship between the rhombohedral axes of the *R*3*m* phase and the face-centered and rhombohedral (dashed line) lattice of the *R*3*c* phase. In the present case the rhombohedral axes of the *R*3*m* phase are nearly orthogonal. The angles between the axes in *R*3*c* phase are approximately 60°.

### B. Raman active modes

**B1**. $Pm\bar{3}m$ phase. The cubic symmetry does not allow Raman active modes, though this rule is virtually always violated in solid solutions.

**B2**. *R*3*m* phase. Raman active modes form the representations $3A_1$+4*E*. The $A_1$ and *E* symmetry modes are roughly split into "transverse" and "longitudinal" modes and, following the notation given in ref. 28 for the tetragonal structure, are labeled as:

$A_1$(*i*TO), $A_1$(*i*LO), *i* = 1, 2, 3, Atoms oscillate mainly along the hexagonal *c*-axis direction

*E*(*j*TO), *E*(*j*LO), *j* = 1, …, 4, Atoms oscillate mainly in the hexagonal *ab*-plane

The indices *i* and *j* increase with increasing Raman shift. In tetragonal PZT the number of Raman active modes is significantly larger than in the isostructural $PbTiO_3$, see refs. [10]. This is expectably the case for rhombohedral PZT. The LO/TO splitting is valid for the *P*4*mm* structure, but not strictly for the *R*3*m* or *R*3*c* phases as each mode involves atomic displacements along the hexagonal *a*- and *c*-axes directions.

**B3.** *R*3*c* phase. The Raman active modes span the representation $4A_1$+9*E* and now *i* = 1,…,4 and *j* = 1,…,9. The *R*3*m*->*R*3*c* transition is signaled by a change in the number of Raman-active modes.

**B4.** The point group symmetry of the rhombohedral phases is 3*m*, the Raman scattering tensors being the same. The Raman scattering tensors for the $A_1$ and *E* symmetry modes are given in Table A1.

Table A1. Raman scattering tensors for the $A_1$ and *E* symmetry modes (point group 3*m*). The coordinates *x*, *y*, or *z* in brackets indicate that the phonon mode is also infrared-active and the mode has the indicated polarization direction.

$A_1(z)$ $\begin{pmatrix} a & 0 & 0 \\ 0 & a & 0 \\ 0 & 0 & b \end{pmatrix}$ $E(y)$ $\begin{pmatrix} c & 0 & 0 \\ 0 & -c & d \\ 0 & d & 0 \end{pmatrix}$ $E(-x)$ $\begin{pmatrix} 0 & -c & -d \\ -c & 0 & 0 \\ -d & 0 & 0 \end{pmatrix}$

These are given in terms of orthogonal axes, the convention being that the *x* and *z* axes are parallel to the hexagonal *a* and *c* axes, respectively.

## C. Raman scattering from an arbitrarily oriented domain

The relationship $R'_{ij} = a_{ip}a_{jq}R_{pq}$ is applied to express the Raman scattering tensors $R_{ij}$ in terms of the new axes *x'*, *y'* and *z'*. The $a_{ip}$ and $a_{jq}$ are the direction cosines. In tetragonal and cubic symmetries the phonons propagating along the crystal axis directions can further be divided into longitudinal and transverse modes. This is no longer rigorously true in the case of rhombohedral symmetry. In the geometry used the phonons do not propagate along the high-symmetry directions, such as crystal axes directions. This means that the phonons are the so-called oblique modes. When giving a symmetry treatment in section C1 we assume that the oblique phonons still have similar atomic displacements as the phonons propagating along the crystal axis directions so that they can be described as possessing either $A_1$- or *E*-symmetry modes. This assumption is justified by noting that in PZT the difference between the *E*- and $A_1$-symmetry modes is quite large, which indicates the crystallographic anisotropy which in turn supports the modes to keep their identity also along general propagation directions. This is just a consequence of the fact that all the phases are essentially pseudosymmetric.

**C1.** Figure 1 shows the domains considered in this study. We assumed that the boundaries are not charged, i.e., we assumed that the difference between the polarization vectors components perpendicular to the domain wall is zero. Table A2 gives the Raman scattering tensors for the domain configurations considered in the text. In a first step a coordinate system corresponding to the conventions for the choice of axes used to define the tensors was defined for each domain. The Raman scattering tensors were then expressed in terms of the *x'*, *y'* and *z'* axes. Table A3 tabulates the Raman scattering intensities for the domains given in Fig. 1 for the measurements geometries used in this study.

Table A2. Raman scattering tensors of the four types of domains (Fig. 1) in terms of the *x'*, *y'* and *z'* axes for the $A_1$ and *E* symmetry modes (point group 3*m*). Also the phonon polarization directions are indicated in terms of the coordinate system *x'*, *y'* and *z'*.

| Domain | $A_1(z)$ | $E(y)$ | $E(-x)$ |
|---|---|---|---|
| 1+<br>$x = -y'$<br>$y = (x' - \sqrt{2}z')/\sqrt{3}$<br>$z = (\sqrt{2}x' + z')/\sqrt{3}$ | $\begin{pmatrix} \frac{a+2b}{3} & 0 & \frac{-\sqrt{2}(a-b)}{3} \\ 0 & a & 0 \\ \frac{-\sqrt{2}(a-b)}{3} & 0 & \frac{2a+b}{3} \end{pmatrix}$ | $\begin{pmatrix} \frac{-c+2\sqrt{2}d}{3} & 0 & \frac{\sqrt{2}c-d}{3} \\ 0 & c & 0 \\ \frac{\sqrt{2}c-d}{3} & 0 & \frac{-2c-2\sqrt{2}d}{3} \end{pmatrix}$ | $\begin{pmatrix} 0 & \frac{c+\sqrt{2}d}{\sqrt{3}} & 0 \\ \frac{c+\sqrt{2}d}{\sqrt{3}} & 0 & \frac{-\sqrt{2}c+d}{\sqrt{3}} \\ 0 & \frac{-\sqrt{2}c+d}{\sqrt{3}} & 0 \end{pmatrix}$ |
| 1-<br>$x = y'$<br>$y = (x' + \sqrt{2}z')/\sqrt{3}$<br>$z = (\sqrt{2}x' - z')/\sqrt{3}$ | $\begin{pmatrix} \frac{a+2b}{3} & 0 & \frac{\sqrt{2}(a-b)}{3} \\ 0 & a & 0 \\ \frac{\sqrt{2}(a-b)}{3} & 0 & \frac{2a+b}{3} \end{pmatrix}$ | $\begin{pmatrix} \frac{-c+2\sqrt{2}d}{3} & 0 & \frac{-\sqrt{2}c+d}{3} \\ 0 & c & 0 \\ \frac{-\sqrt{2}c+d}{3} & 0 & \frac{-2c-2\sqrt{2}d}{3} \end{pmatrix}$ | $\begin{pmatrix} 0 & \frac{-c-\sqrt{2}d}{\sqrt{3}} & 0 \\ \frac{-c-\sqrt{2}d}{\sqrt{3}} & 0 & \frac{-\sqrt{2}c+d}{\sqrt{3}} \\ 0 & \frac{-\sqrt{2}c+d}{\sqrt{3}} & 0 \end{pmatrix}$ |
| 2+<br>$x = x'$<br>$y = (y' + \sqrt{2}z')/\sqrt{3}$<br>$z = (-\sqrt{2}y' + z')/\sqrt{3}$ | $\begin{pmatrix} a & 0 & 0 \\ 0 & \frac{a+2b}{3} & \frac{\sqrt{2}(a-b)}{3} \\ 0 & \frac{\sqrt{2}(a-b)}{3} & \frac{2a+b}{3} \end{pmatrix}$ | $\begin{pmatrix} c & 0 & 0 \\ 0 & \frac{-c-2\sqrt{2}d}{3} & \frac{-\sqrt{2}c-d}{3} \\ 0 & \frac{-\sqrt{2}c-d}{3} & \frac{-2c+2\sqrt{2}d}{3} \end{pmatrix}$ | $\begin{pmatrix} 0 & \frac{-c+\sqrt{2}d}{\sqrt{3}} & \frac{-\sqrt{2}c-d}{\sqrt{3}} \\ \frac{-c+\sqrt{2}d}{\sqrt{3}} & 0 & 0 \\ \frac{-\sqrt{2}c-d}{\sqrt{3}} & 0 & 0 \end{pmatrix}$ |
| 2-<br>$x = x'$<br>$y = (-y' + \sqrt{2}z')/\sqrt{3}$<br>$z = (-\sqrt{2}y' - z')/\sqrt{3}$ | $\begin{pmatrix} a & 0 & 0 \\ 0 & \frac{a+2b}{3} & \frac{-\sqrt{2}(a-b)}{3} \\ 0 & \frac{-\sqrt{2}(a-b)}{3} & \frac{2a+b}{3} \end{pmatrix}$ | $\begin{pmatrix} c & 0 & 0 \\ 0 & \frac{-c+2\sqrt{2}d}{3} & \frac{\sqrt{2}c-d}{3} \\ 0 & \frac{\sqrt{2}c-d}{3} & \frac{-2c-2\sqrt{2}d}{3} \end{pmatrix}$ | $\begin{pmatrix} 0 & \frac{c+\sqrt{2}d}{\sqrt{3}} & \frac{-\sqrt{2}c+d}{\sqrt{3}} \\ \frac{c+\sqrt{2}d}{\sqrt{3}} & 0 & 0 \\ \frac{-\sqrt{2}c+d}{\sqrt{3}} & 0 & 0 \end{pmatrix}$ |

Table A3. Raman scattering intensity factors expected for the four types of domains considered in this study. The intensity is directly proportional to these factors. The Table indicates that the measured geometries do not distinguish between domains 1+ and 1-. In principle, the $z'(x'x')\overline{z'}$ and $z'(y'y')\overline{z'}$ geometries reveal both $A_1$ and $E$ symmetry modes, whereas the $z'(x'y')\overline{z'}$ geometry shows only $E$-symmetry modes. The phonon propagation direction is along the rhombohedral $c$-axis direction and thus the modes have both transverse and longitudinal features. Coordinates $x$, $y$ and $z$ are domain-specific and are given in Table A2.

| Domain | Geometry | Intensity factor |
|---|---|---|
| 1+, 1- | $z'(x'x')\overline{z'}$ | $A_1(z)$: $(a+2b)^2/9$, $E(y)$: $\left(-c+2\sqrt{2}d\right)^2/9$, $E(-x)$: $0$ |
| | $z'(x'y')\overline{z'}$ | $A_1(z)$: $0$, $E(y)$: $0$, $E(-x)$: $\left(c+\sqrt{2}d\right)^2/3$ |
| | $z'(y'y')\overline{z'}$ | $A_1(z)$: $a^2$, $E(y)$: $c^2$, $E(-x)$: $0$ |
| 2+ | $z'(x'x')\overline{z'}$ | $A_1(z)$: $a^2$, $E(y)$: $c^2$, $E(-x)$: $0$ |
| | $z'(x'y')\overline{z'}$ | $A_1(z)$: $0$, $E(y)$: $0$, $E(-x)$: $\left(-c+\sqrt{2}d\right)^2/3$ |
| | $z'(y'y')\overline{z'}$ | $A_1(z)$: $(a+2b)^2/9$, $E(y)$: $\left(c+2\sqrt{2}d\right)^2/9$, $E(-x)$: $0$ |
| 2- | $z'(x'x')\overline{z'}$ | $A_1(z)$: $a^2$, $E(y)$: $c^2$, $E(-x)$: $0$ |
| | $z'(x'y')\overline{z'}$ | $A_1(z)$: $0$, $E(y)$: $0$, $E(-x)$: $\left(c+\sqrt{2}d\right)^2/3$ |
| | $z'(y'y')\overline{z'}$ | $A_1(z)$: $(a+2b)^2/9$, $E(y)$: $\left(-c+2\sqrt{2}d\right)^2/9$, $E(-x)$: $0$ |

## D. Raman mode assignment

For a qualitative mode assignment we utilize our earlier density functional theory computations dedicated for addressing the properties of $PbTiO_3$ under hydrostatic pressure. The results given below were obtained in the context of the studies [27,29]. We note that the ground state of $PbTiO_3$ is the *P*4*mm* structure, as computations suggest: no phonon instabilities were observed at Brillouin zone, see Fig. A10. Thus, besides the usual pseudo-potential tests also considerable amount of computational work was conducted to ensure the stability issue. Stability is a crucial aspect for phonon frequency computation. We note that *ab initio* computation of phonon frequencies for a disordered PZT is still a formidable task as the required supercells are well beyond the present computational capacity.

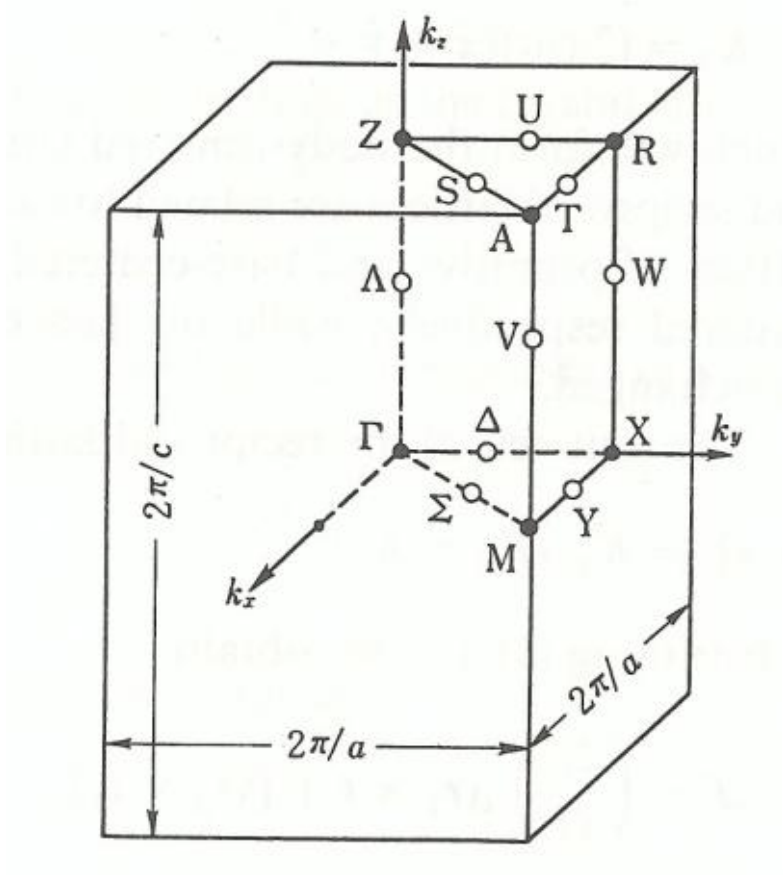

| **k** | Irrep | Frequencies ($cm^{-1}$) |
|---|---|---|
| *Z* | $A_1$ | 109, 219, 452, 823 |
| | $B_1$ | 294 |
| | *E* | 52, 160, 197, 278, 497 |
| *X* | $A_1$ | 69, 249, 289, 312, 497 |
| | $A_2$ | 142, 256, 476 |
| | $B_1$ | 57, 321 |
| | $B_2$ | 107, 181, 332, 609, 705 |
| *M* | $A_1$ | 73, 469 |
| | $A_2$ | 460 |
| | $B_1$ | 111 |
| | $B_2$ | 244, 618, 767 |
| | *E* | 62, 221, 300, 412 |
| *R* | $A_1$ | 68, 108, 280, 421, 424 |
| | $A_2$ | 163, 288, 456 |
| | $B_1$ | 57, 334 |
| | $B_2$ | 62, 217, 319, 615, 802 |
| *A* | $A_1$ | 89, 414 |
| | $A_2$ | 458 |
| | $B_1$ | 107 |
| | $B_2$ | 233, 653, 820 |
| | *E* | 72, 130, 340, 401 |

| *Γ*-point mode | Computed Frequency ($cm^{-1}$) | Experimental Frequency ($cm^{-1}$) |
|---|---|---|
| *E*(1TO) | 82.5 | 87.5 |
| *E*(1LO) | 124.1 | 128.0 |
| $A_1$(1TO) | 147.6 | 148.5 |
| $A_1$(1LO) | 190.0 | 194.0 |
| *E*(2TO) | 191.9 | 218.5 |
| *E*(2LO) | 273.7 | No datumn |
| *E*(3TO) | 274.2 | 289.0 |
| $B_1$ | 281.0 | 289.0 |
| $A_1$(2TO) | 344.5 | 359.5 |
| *E*(3LO) | 418.9 | 440.5 |
| $A_1$(2LO) | 443.8 | 465.0 |
| *E*(4TO) | 507.0 | 505.0 |
| $A_1$(3TO) | 637.5 | 647.0 |
| *E*(4LO) | 683.9 | 687.0 |
| $A_1$(3LO) | 780.3 | 795.0 |

Fig. A10. Left: Brillouin Zone (BZ) of the *P*4*mm* structure of $PbTiO_3$. Righ: Phonon frequencies at the high-symmetry points of the BZ. No phonon instabilities were observed and the BZ center modes frequencies are consistent with the experimental values [30].

Computations predict that under hydrostatic pressure (critical value approximately 9GPa) the *R*3*c* phase is stabilized in $PbTiO_3$ [27,29]. The phase is structurally the same as the idealized *R*3*c* phase of $Pb(Zr_{0.80}Ti_{0.20})O_3$. Thus, the phonon properties are expected to be qualitatively similar. Though the *R*3*m* phase is not stable at any of studied hydrostatic pressure values, values for a hypothetical *R*3*m* phase are given in order to demonstrate the difference between the two phases. The phonon frequencies are also plotted against hydrostatic pressure (different oxygen octahedral tilts and bond lengths) in order to show that the low-frequency spectra of *R*3*c* and *R*3*m* remain very different in the whole pressure range: pressure does not introduce any accidental degeneracies in the low-frequency range of the *R*3*c* phase.

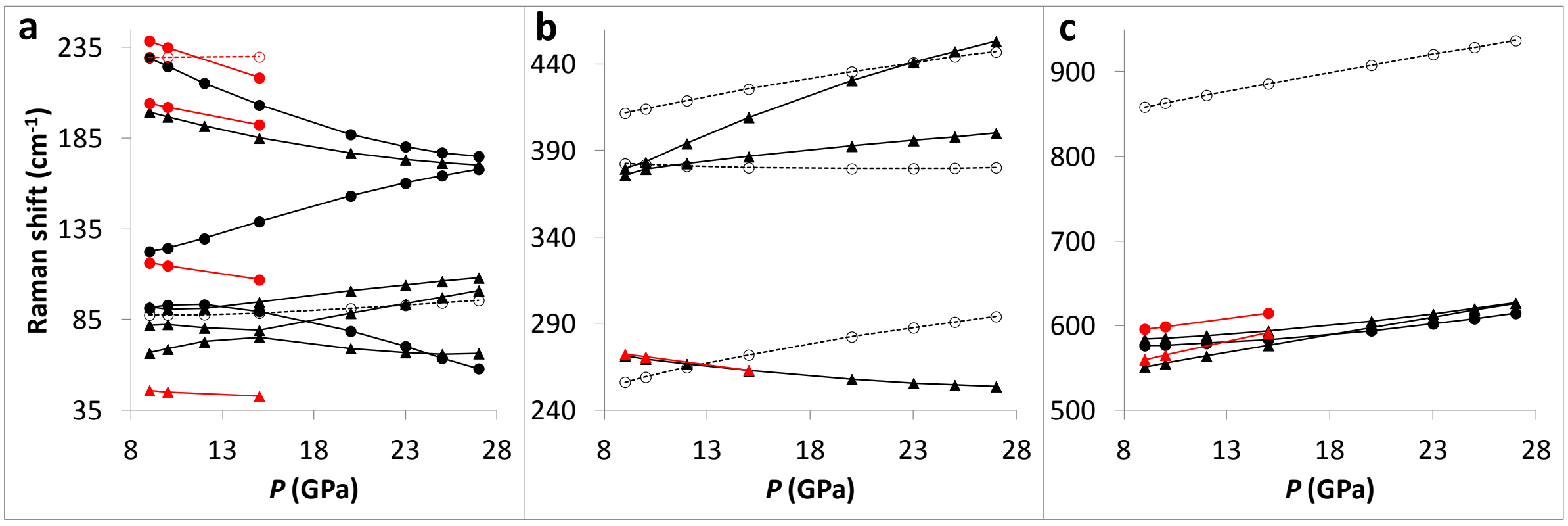

Fig. A11. Computed (a) low-, (b) intermediate and (c) high-frequency Brillouin zone center mode positions. No LO/TO splitting was taken into account. For LO/TO splitting see Tables A4 and A5. The black and red data points are for the *R*3*c* and *R*3*m* structures, respectively. Filled spheres, open spheres and triangles correspond to the $A_1$, $A_2$ and *E*-symmetry modes, respectively. $A_2$ modes are not Raman active, though they could be activated by disorder in PZT.

Table A4. Computed TO and LO Brillouin zone center mode frequencies of the *R*3m phase at 9 GPa. The electric field connected to the longitudinal phonon modes tabulated on the left-and right-hand sides are along the *z* and *x*-axis directions, respectively. The integer in front of TO/LO is a running index for bookkeeping purposes and the modes possessing the same label in this Table and Table A5 are not comparable.

| Mode | Raman shift ($cm^{-1}$) | Mode | Raman shift ($cm^{-1}$) |
|---|---|---|---|
| *E*(1TO) | 44.2 | *E*(1LO) | 140.7 |
| *E*(2TO) | 203.6 | *E*(2LO) | 271.9 |
| *E*(3TO) | 271.9 | *E*(3LO) | 416.5 |
| *E*(4TO) | 559.5 | *E*(4LO) | 752.7 |
| $A_1$(1LO) | 150.9 | $A_1$(1TO) | 115.4 |
| $A_1$(2LO) | 442.3 | $A_1$(2TO) | 238.0 |
| $A_1$(3LO) | 747.1 | $A_1$(3TO) | 595.7 |
| $A_2$ | 229.3 | $A_2$ | 229.3 |

Table A5. Computed TO and LO Brillouin zone center mode frequencies of the *R*3*c* phase at 9 GPa. The electric field connected to the longitudinal phonon modes tabulated on the left-and right-hand sides are along the *z* and *x*-axis directions, respectively.

| Mode | Raman shift ($cm^{-1}$) | Mode | Raman shift ($cm^{-1}$) |
|---|---|---|---|
| *E*(1TO) | 67.1 | *E*(1LO) | 81.3 |
| *E*(2TO) | 81.3 | *E*(2LO) | 87.4 |
| *E*(3TO) | 88.9 | *E*(3LO) | 144.6 |
| *E*(4TO) | 197.1 | *E*(4LO) | 268.8 |
| *E*(5TO) | 269.0 | *E*(5LO) | 379.6 |
| *E*(6TO) | 379.7 | *E*(6LO) | 381.6 |
| *E*(7TO) | 383.1 | *E*(7LO) | 423.8 |
| *E*(8TO) | 554.3 | *E*(8LO) | 584.0 |
| *E*(9TO) | 584.4 | *E*(9LO) | 747.0 |
| $A_1$(1LO) | 110.1 | $A_1$(1TO) | 90.9 |
| $A_1$(2LO) | 149.5 | $A_1$(2TO) | 123.2 |
| $A_1$(3LO) | 432.5 | $A_1$(3TO) | 224.7 |
| $A_1$(4LO) | 739.6 | $A_1$(4TO) | 576.0 |
| $A_2$(1) | 86.9 | $A_2$(1) | 86.9 |
| $A_2$(2) | 258.9 | $A_2$(2) | 258.9 |
| $A_2$(3) | 382.2 | $A_2$(3) | 382.2 |
| $A_2$(4) | 413.9 | $A_2$(4) | 413.9 |
| $A_2$(5) | 862.5 | $A_2$(5) | 862.5 |

**Appendix II: Model for a thermal strain**

Spatial composition variation in PZT is an inherent property: despite the rather large size difference between the $Ti^{4+}$ and $Zr^{4+}$ cations, no long-range ordering takes place and a statistical distribution is a good approximation. In the sample synthesis it is likely that even larger variations occur. We have recently developed a method for treating such materials [31]. In crystals the method divides the material into thin slices in such a way that the slices are smoothly joined together. The method involves breaking the translational symmetry and allows a spatially varying composition and bond-length distribution.
To describe the observed high-temperature Raman peaks (Figs. 1, 6 and 7) we divide the material into slices, whose composition and thickness is allowed to vary, Fig. A12.

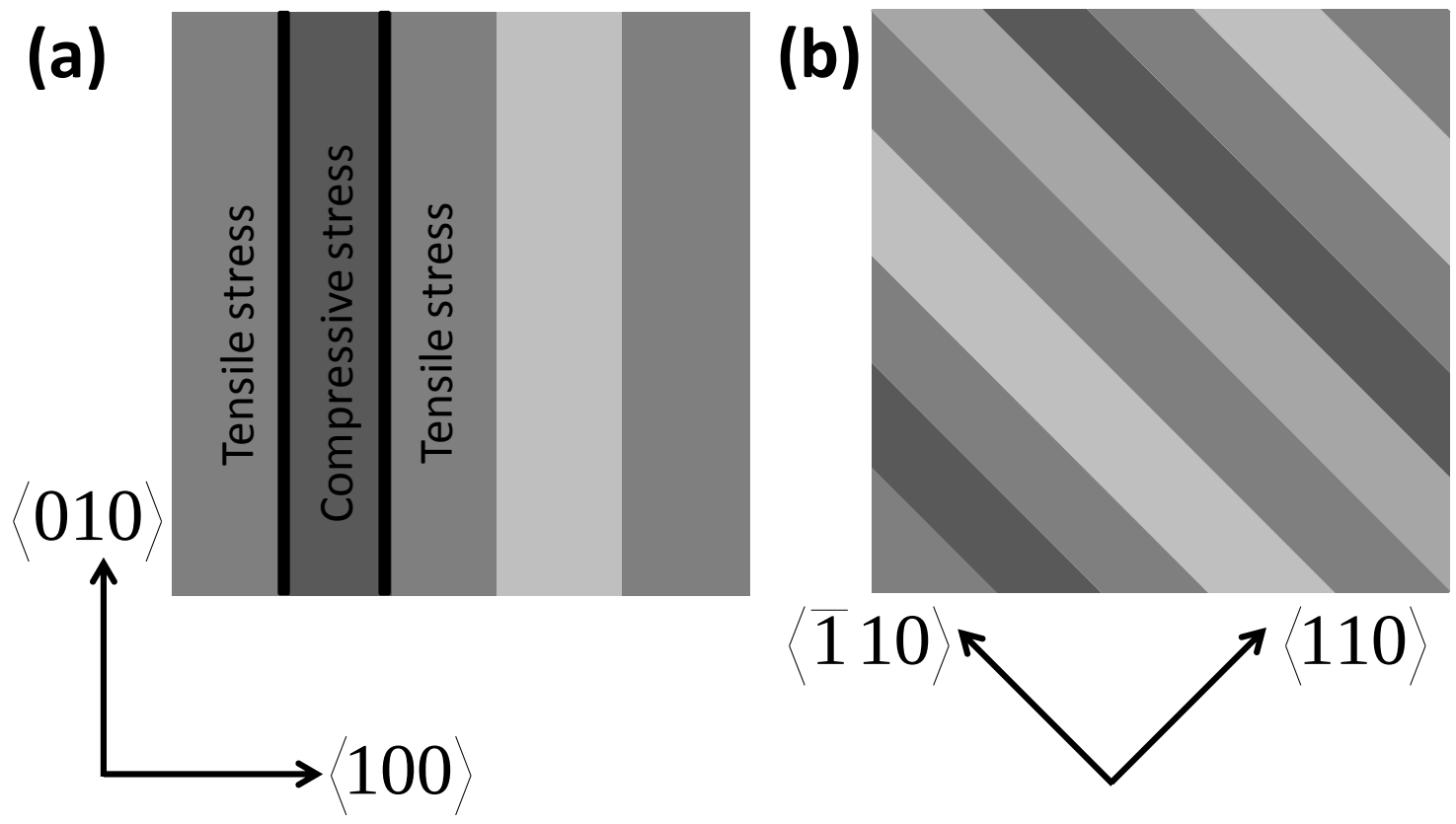

Fig. A12. A schematic model for a spatial composition-variation-induced strain in PZT crystal. The crystal is divided into slices colored by different tones of gray: the darker the color the larger the Zr-concentration. The color indicates the average slice composition, which in turn causes different slice thicknesses and different thermal expansions. Panel (a) illustrates the interpretation for Figs. 1 and 6, and panel (b) illustrates the interpretation for Fig. 7. Panel (a) also depicts the formation of interface (black pillars) surrounding the Zr-rich slice.

When the crystal is heated, slices having larger (smaller) thermal expansion coefficient with respect to their neighbors are under compressive (tensile) stress. Eventually large stress results in symmetry changes and activation of Raman modes forbidden in the ideal cubic structure. It is worth stressing that the high-temperature Raman modes were observed only in selected areas possessing stripes which probably correspond to the largest composition gradients across the boundaries. A more accurate treatment would require the treatment of all slices in order to determine the stress state of each slice as the slices are all coupled.

It is difficult to assign a specific symmetry merely based on a few peaks as the scattering takes place from a structure possessing several slices. To illustrate how Raman scattering could take place in simplest terms we compiled Table A6. Fig. 4d shows that the peaks at around 84, 138 and 273 $cm^{-1}$ are clearly observed with crossed polarizers, and nearly vanish with parallel polarizers, and thus one could assign them to the three $T_2$ modes of the point group $T_d$. However, the intensity of the band between 400 and 600 $cm^{-1}$ is stronger when the measurement is conducted with parallel polarizers. This suggests that the symmetry is probably lower than $P\bar{4}3m$.
We note that a symmetry lowering to $R\bar{3}m$ is consistent with the behavior observed in Fig. 7: low-frequency features are observed with crossed and parallel polarizers (i.e., they are not totally symmetric modes), whereas the band between 400 and 600 $cm^{-1}$ vanishes with crossed polarizers. Table A3 indicates that it could consist of $A_{1g}$ and/or $E_g$-symmetry modes (note that the analysis given in Table A3 is valid for the point group $\bar{3}m$).
For practical applications a deposition of thin-film layers with controlled Zr/Ti concentration gradient is a way to make homogeneous structures with symmetries different from the ones reported for bulk materials.

Table A6. Type I subgroups of the space group $Pm\bar{3}m$ and the corresponding irreducible representation of the Brillouin zone center modes. Letters A, IR and R in braces indicate acoustic infrared-active and Raman-active modes, respectively.

| Subgroup | Irreducible representation |
|---|---|
| $P\bar{4}3m$ | $T_2$(A)+3$T_2$(IR,R)+ $T_1$ |
| $P432$ | $T_1$(A)+3$T_1$(IR)+$T_2$(R) |
| $Pm\bar{3}$ | $T_u$(A)+4$T_u$ |
| $R\bar{3}m$ | $A_{2u}$(A)+ $E_u$(A)+$A_{1g}$(R)+$E_g$(R)+2$A_{2u}$(IR)+2$E_u$(IR) |